\documentclass{iau}
\usepackage{graphicx}

\newcommand{\atlas}{ATLAS$^{\rm 3D}$}

\title[AGN Feedback in NGC~1266] 
{AGN Feedback Driven Molecular Outflow in NGC~1266}

\author[K. Alatalo and the ATLAS$^{\rm 3D}$ team]   
{K.\ Alatalo$^1$, K.\ E.\ Nyland$^2$, G.\ Graves$^1$, S.\ Deustua$^3$, J. Wrobel,$^4$ L.~M. Young$^2$, T.~A. Davis$^5$, M. Bureau$^6$, E. Bayet,$^6$ L.\ Blitz,$^2$ M.\ Bois,$^7$ F.\ Bournaud,$^8$, M. Cappellari,$^6$ R.\ L.\ Davies,$^6$ P.~T.\ de~Zeeuw,$^{5,14}$ E.\ Emsellem,$^{5,17}$ S.\ Khochfar,$^9$ D.\ Krajnovic,$^5$ H.\ Kuntschner,$^5$ S. Mart\'in,$^{10}$ R.\ M.\ McDermid,$^{11}$ R.\ Morganti,$^{12}$ T. Naab,$^{9}$ T. Oosterloo,$^{12}$ M.\ Sarzi,$^{13}$ N. Scott,$^{15}$ P.\ Serra$^{13}$ and A.\ Weijmans$^{16}$}

\affiliation{
$^1$University of California, Berkeley, USA;
$^2$New Mexico Tech, Socorro, USA;
$^3$Space Telescope Science Institute, Baltimore, USA;
$^4$NRAO, Socorro, USA;
$^5$ESO, Garching, Germany;
$^6$University of Oxford, UK;
$^7$Observatoire de Paris, France;
$^8$Universit\'e Paris Diderot, France;
$^9$MPI for Extraterrestrial Physics, Garching, Germany;
$^{10}$ESO, Santiago, Chile;
$^{11}$Gemini Observatory, Hilo, USA;
$^{12}$ASTRON, Dwingeloo, The Netherlands;
$^{13}$University of Hertfordshire, Hatfield, UK;
$^{14}$Leiden University, The Netherlands;
$^{15}$Swinburne University, Australia;
$^{16}$University of Toronto, Canada;
$^{17}$Universit\'{e} de Lyon, France}

\pubyear{2012}
\volume{290}  
\jname{Feeding compact objects: Accretion on all scales}
\editors{C.M. Zhang, T. Belloni, M. M\'endez \& S.N. Zhang, eds.}

\begin{document}

\maketitle

\begin{abstract}
NGC~1266 is a nearby field galaxy observed as part of the \atlas\ survey (Cappellari et al. 2011).  NGC~1266 has been shown to host a compact ($< 200$ pc) molecular disk and a mass-loaded molecular outflow driven by the AGN (Alatalo et al. 2011).  Very Long Basline Array (VLBA) observations at 1.65 GHz revealed a compact (diameter $< 1.2$ pc), high brightness temperature continuum source most consistent with a low-level AGN origin.  The VLBA continuum source is positioned at the center of the molecular disk and may be responsible for the expulsion of molecular gas in NGC 1266.  Thus, the candidate AGN-driven molecular outflow in NGC 1266 supports the picture in which AGNs do play a significant role in the quenching of star formation and ultimately the evolution of the red sequence of galaxies.
\end{abstract}

\firstsection
\vspace{-3mm}
\section{Introduction}
The present-day galaxy population has a bimodal color distribution, with a genuine lack of green valley galaxies (Strateva et al. 2001; Baldry et al. 2004).  This dearth of green valley galaxies, coupled with the increase in quiescent galaxies since $z \sim 1$, implies that galaxies quench star formation (SF) and transition rapidly from blue to red optical colors (Bell et al. 2004; Faber et al. 2007).  Thus far, it is not clear if, or how, AGN feedback plays a role in this rapid quenching (Springel, di Matteo \& Hernquist 2005; Hopkins et al. 2005), though there is some circumstantial evidence that supports the AGN-driven SF quenching scenario.  In particular, Schawinski et al.\ (2007) have found that AGNs predominantly reside in green valley galaxies, but direct evidence (e.g., a galaxy where the AGN is clearly responsible for the expulsion of starforming material) has been scarce.  However, the recent discovery of a massive molecular outflow likely driven by an AGN in the early-type galaxy NGC~1266 may be a rare example of a galaxy transitioning from the blue to red sequence.  NGC 1266 is thus a promising local laboratory for studying AGN-driven SF quenching.

\vspace{-6mm}
\section{Overview}
\begin{figure}[h]
\begin{center}
\includegraphics[width=5.4in]{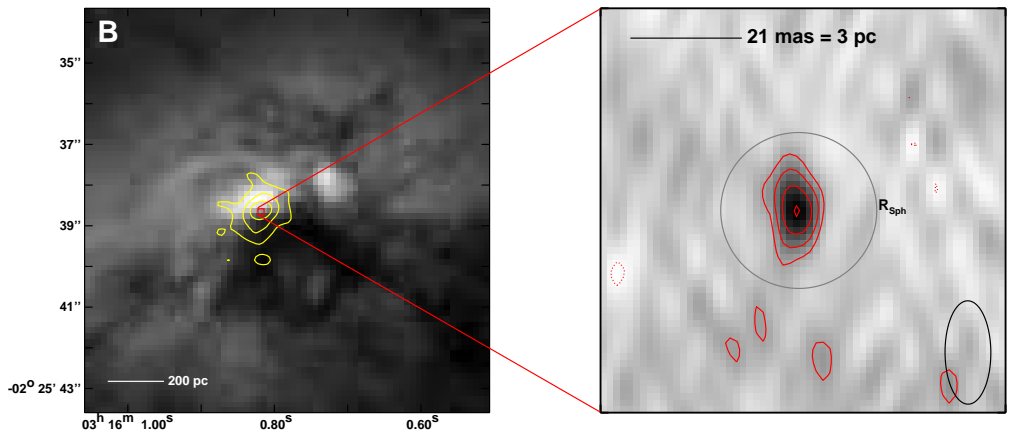}
\vspace{-7mm}
\caption{{\bf (Left):} NGC 1266 Hubble Space Telescope $B$-band greyscale imaged.  CARMA CO(1--0) $v_{\rm sys}$ contours are overlaid in yellow at levels of [0.25, 0.5, 0.75] of the peak CO emission ($\gtrsim 1937$ K km s$^{-1}$).  The box shows the field-of-view of the VLBA image to the right.  {\bf (Right):} NGC~1266 1.65 GHz continuum image with contours.  The VLBA beam is the black ellipse in the lower right corner and has a major axis diameter of 9.75 mas (1.14 pc).  The relative contour levels are [-3, 3, 6, 10, 14] and the unit of the contour level is 42 $\mu$Jy beam$^{-1}$.  The AGN is detected to 14$\sigma$ significance, and the VLBA does indeed probe the sphere of influence ($R_{\rm Sph}$), but does not appear to resolve the point source.}
\label{fig1}
\end{center}
\end{figure}

NGC~1266 is a nearby ($D \approx 29.9$ Mpc) field early-type galaxy that CO observations have shown to harbor a compact molecular disk ($\Sigma_{\rm gas} > 10^4~M_\odot$ pc$^{-2}$) and a mass-loaded ($M_{\rm gas} > 10^9~M_\odot$; $\dot{M} > 13~M_\odot$ yr$^{-1}$) molecular outflow (Alatalo et al. 2011).  Alatalo et al.\ (2011) showed that the molecular gas could not be driven by SF alone and concluded that an AGN was the most likely culprit.  Very Large Array (VLA) observations revealed a compact radio core coincident with the weak detection of hard X-rays with {\em Chandra} (Alatalo et al. 2011).  Subsequent VLBA observations confirmed the presence of a compact, high brightness temperature ($T_b > 10^7$ K) radio core (Fig. \ref{fig1}), further supporting existing evidence that NGC~1266 hosts a low-level AGN.  The VLBA source is positioned at the center of the molecular disk (Fig. \ref{fig1}a, yellow contours), where the H$_2$ column density is at least $6\times10^{24}$ cm$^{-2}$.  Given the high column density of molcular gas directly atop the AGN, it is unsurprising that little hard X-ray emission was detected (Alatalo et al. 2011) since this object may be mildly compton thick.  VLBA observations have provided strong supporting evidence for the presence of an AGN in the center of NGC 1266, evidence directly implicating the AGN in the outflow awaits .  Future High Sensitivity Array observations will provide improved imaging capable of revealing the signature of a radio outflow at the launch point of the outflowing molecular gas and would provide direct evidence of an AGN-driven molecular outflow in the local universe.

\end{document}